\documentclass[aps,twocolumn,showpacs]{revtex4}
\usepackage{graphicx}
\usepackage{dcolumn}
\usepackage{bm}

\begin{document}

\title{Deceleration mechanism of Fermi acceleration in a time-dependent stadium billiard}

\author{Andr\'e L.\ P.\ Livorati$^1$, Alexander Loskutov$^{1,2}$ and
Edson D.\ Leonel$^1$}
\affiliation{$^1$ Departamento de Estat\'istica, Matem\'atica Aplicada e
Computa\c c\~ao -- Univ Estadual Paulista\\ Av.24A, 1515 -- Bela Vista
-- CEP: 13506-900 -- Rio Claro -- SP -- Brazil\\
$^{2}$Physics Faculty, Moscow State University, 119992 Moscow, Russia}


\begin{abstract}
The dynamics of a time-dependent stadium-like billiard are studied by a four dimensional nonlinear mapping. We have shown that even without any dissipation, the particle experiences a decrease on its velocity. Such condition is related with a critical resonance velocity, where if the initial velocity has a higher value than the resonant one, we can observe Fermi acceleration, however, if the initial velocity has a initial value smaller than the critical one, the particle is temporarily trapped surrounding the stability islands, in a stickiness regime. We believe that this sticky orbits can act as deceleration mechanism for Fermi Acceleration.
\end{abstract}

\maketitle

Problems involving particles colliding (with or without interaction among themselves) into a closed domain, are well known in the literature as billiard problems.
The laws of reflection of the point particle (billiard ball) are the same for the light's specular reflection, i. e. while the normal component changes its sign on the collision point, the tangent one is preserved and the particle is reflected elastically \cite{ref1}.
The study of billiard problems had begun with Birkhoff \cite{ref2}, but it was Sinai \cite{ref3} that gave us the necessary mathematical formulation for a new rigorous analysis.
Nowadays, billiard problems can be found in several fields of physics and other sciences, like optics \cite{ref4}, microwaves \cite{ref5}, quantum dots \cite{ref6}, ultra-cold atoms trapped in a laser potential \cite{ref7,ref8}, among others.\

If the billiard boundaries are time-dependent, we may found an interesting phenomenon called Fermi Acceleration (FA). Introduced for the first time in 1949 by Italian physicist Enrico Fermi \cite{ref9}, FA is basically the unlimited energy growth for a classical particle suffering elastic collisions with a time-dependent boundary.
For one dimensional billiards we have some examples of this unlimited growth of energy \cite{p1,ref10,ref11,ref12,ref13}.
In two dimensional billiards, in which the dynamics are more complicated, this phenomenon has already been studied in different models \cite{ref14,ref15,ref17,ref18,ref19,ref20}. One of the main questions about FA in perturbed systems is whether the energy of the particle can grow to infinity. The answer is far away from trivial, and it depends on the kind of perturbation on the boundary as well the boundaries geometry. The Loskutov-Ryabov-Akinshin (LRA) conjecture \cite{ref21} tell us that if a billiard with fixed boundary presents chaotic behavior, Fermi acceleration is exhibited, when perturbation on the boundary is introduced. However, F.Lenz et. al \cite{ref22}, shown that even for an elliptical billiard, which has regular dynamics in the static boundary case, exhibits FA when time-dependence is introduced. In this particular case, FA production mechanism is due orbits that crosses the stochastic layer, changing their dynamics from librator to rotator, or vice-versa. And recently, Leonel and Bunimovich \cite{ref23}, extended the LRA conjecture, by showing that only the presence of heteroclinc orbits in the static phase space, is sufficient condition for the system present FA when time-dependence is introduced.

There are some methods to suppress FA, like introducing dissipation via inelastic collisions \cite{ref13,ref14,ref23}, or through a drag force \cite{ref20}. But, in this letter, we found that orbits in a stickiness regime \cite{ref26,ref27} characterized by a resonance, acts as a deceleration mechanism of the particle average velocity. This stickiness phenomenon  has already been studied in some billiards \cite{ref17,ref97,ref98} with different approaches, but in this letter we intent to show how this stickiness, can be used as a deceleration mechanism for FA, even with no dissipation introduced at all.\ 

In this letter, a stadium-like billiard is studied considering the dynamics proposed by Loskutov et. al. \cite{ref21}, which describes the billiard dynamics by a nonlinear map and considered the fixed boundary approximation, i. e., the boundary is considered as fixed, but when the particle collides with it, they exchange momentum, as if the boundary were moving through a periodic time perturbation. A resonance critical value for the particle's velocity, depending on the geometrical control parameters of the billiard is found by linearizing the unperturbed map. Our results show that depending on the initial velocity of the particle we can found FA for initial velocities above this critical value. For values of initial velocities lower than the critical one, we observe a decreasing on velocity even with no dissipation. This decreasing is related by some orbits that rotates around the fixed points due a resonance on the boundaries perturbation, in a regime of sticky orbits. As a consequence, the particle average velocity experiences a deceleration.\

\begin{figure}[h]
\centering
\includegraphics[width=4.5cm]{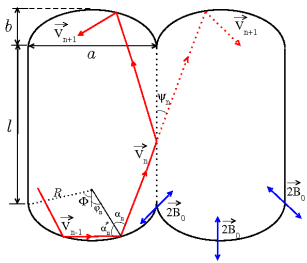}
\caption{(color online). The unfolding method, the geometrical control parameters and a trajectory with indirect collisions.}
\label{fig1}
\end{figure}

The model under study consists of a classical particle suffering elastic collisions inside a stadium-like billiard with focusing components that are periodically time-dependent according to $B(t)=B_0 \cos(wt)$, where $w$ is the frequency of oscillation and $B_0$ is the amplitude of oscillation. The geometrical control parameters are described in Fig.\ref{fig1}. Suppose that focusing components, which are symmetric about the vertical billiard axis, had a radius $R$ with the angle measure $2\Phi$. Looking at the Fig.\ref{fig1}, one can geometrically obtain the billiard parameters as $R=(a^2 + 4b^2)/8b$ and $\Phi=\arcsin(a/2R)$.\ 

The initial angular conditions are $(\alpha,\varphi)$. Just for notation, all the variables with a star index, are measured just before the collision point. We assume that $V_n$ is the particle velocity and $t_n$ is the time of the $n^{th}$ collision, and of course, at initial time $t_0 = 0$, the particle belongs to the focusing component and the velocity vector directs towards to the billiard table. In order to describes the dynamics of this billiard, we should consider two different cases. In the first one, the particle collides with a boundary component and the subsequent collision is with the same focusing component. The second case, the particle collides with the focusing component, and in the next collision, the particle hits opposite boundary component, so in this case we use the unfolding method \cite{ref1,ref21} to describes its dynamics.
For both cases, the recurrence relation for the velocity and the angle $\alpha_n$ are the same. Making a vectorial arrangement for $\vec{V}_{n+1}$ and $\vec{V}_n$ with the angles $\alpha_n$ and ${\alpha^{*}_n}$ and applying the cosine law for the velocity vectors $\vec{V}_n$, $\vec{V}_{n+1}$ and $\vec{B}_n$, according to Fig.\ref{fig1} we found:
\begin{equation}
\alpha_n = arcsin\left({V_n\over V_{n+1}}\sin({\alpha^{*}_n})\right).
\label{eq1}
\end{equation}
\begin{equation}
V_{n+1} = \sqrt{{V_n}^2 + 4{B_n}^2 + 4V_nB_n\cos({\alpha^*_n})}~,
\label{eq2}
\end{equation}
where $B_n$ is the boundary velocity vector in the time $t_n$.\

\begin{figure}[h!]
\centering
\includegraphics[width=4.5cm]{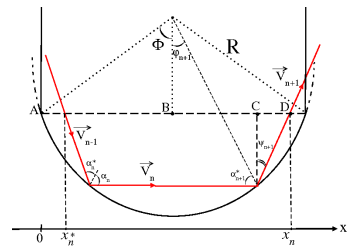}
\caption{(color online). Some dynamical variables and a trajectory with successive collisions.}
\label{fig2}
\end{figure}

Let us now introduce the dynamical equations for the first case, i.e., the successive collisions with the same focusing component. This kind of collisions will only happen when $\Phi \ge \mid{\varphi_{n+1}}\mid$. According to the figure \ref{fig2} and the specular reflection law, we can obtain these relations:
\begin{eqnarray}
\alpha^*_{n+1} &=& \alpha_n~ \nonumber \\
\varphi_{n+1} &=& \varphi_n +\pi -2\alpha_n~~ (mod~ 2\pi)~. \nonumber \\
t_{n+1} &=& t_n + {2R\cos(\alpha_n)\over V_n}~. \nonumber \\
\label{eq2}
\end{eqnarray}

Now, we should consider the case where the particle has its next collision with the opposite focusing component. For this case, it is necessary that $\Phi < \mid{\varphi_{n+1}}\mid$, and, we will introduce some more dynamical variables, as $\psi$, that is the angle between the particle trajectory and the vertical line at the collision point, and $x_n$, that is the projection of the particle position under the horizontal axis.\

Looking at the figure \ref{fig2} we can geometrically obtain the value of the angle $\psi_n=\alpha_n-\varphi_n$. We also can see that $x_n$ is the sum of the line segments $\overline{AB}+\overline{BC}+\overline{CD}$. Taking into account the value of $\psi_n$, and, after some geometrical algebra, we obtain $x_n={R\over\cos(\psi_n)}[\sin(\alpha_n)+\sin(\Phi-\psi_n)]$. The recurrence relation between $x_n$ and $x_{n+1}$ is given by the unfolding method, described in Fig.\ref{fig1}, as $x_{n+1}=x_n+l\tan(\psi_n)$.\

Now let us find the equations for the angular dynamical variables and time. If we invert the particle motion, i.e. consider the reverse direction of the billiard particle, then the expression that furnishes us the value of $x_n$ is also inverted and the angle $\alpha_n$, became $\alpha^{*}_n$. Resolving it with respect to $\alpha^{*}_n$, taking into account that this angle is changed in the opposite direction than $\alpha_n$, and the angle $\varphi_n$ should have the reversed sign, we now can get the value of the incident angle $\alpha^{*}_n$, that will become $\alpha^{*}_{n+1}$ when we re-inverted the particle motion.
The values of $\varphi_{n+1}$ and the time $t_{n+1}$ can be obtained by easy geometrical considerations on Fig.\ref{fig1}. Thus, we obtain the mapping the case where collisions with the opposite focusing component happen.

\begin{eqnarray}
\alpha^{*}_{n+1}&=&\arcsin[\sin(\psi_n+\Phi)-x_{n+1}\cos(\psi_n)/R]~.  \nonumber  \\
\varphi_{n+1}&=&\psi_n-\alpha^*_{n+1}~. \nonumber \\
t_{n+1}&=&t_n+{R[\cos(\varphi_n)+\cos(\varphi_{n+1})-2\cos(\Phi)]+l\over V_n\cos(\psi_n)}~. \nonumber \\
\label{eq3}
\end{eqnarray}

When we linearize the unperturbed map \cite{ref21,ref24}, we may obtain, according the action-angle variables, a rotation number as $\sigma=\arccos(1-8bl/a^2\cos^2(\psi^{*}_n))$, where $\psi^{*}_n=\arctan(ma/l)$ is the fixed point, where $m\ge 1$, is the number of mirrored stadiums in the unfolding method. Considering a trajectory where the particle moves around some stable fixed point, the time between two sequential collisions is given by $\tau \approx l/\cos(\psi^{*}_n)V_n$. Thus, the rotation period is $T_{rot}=2\pi\tau/\sigma$. If the rotation period is equal to the period of external perturbation $T_{ext}=2\pi/\omega$, we observe a resonance between rotation and boundary oscillations. This resonant velocity is given by:
\begin{equation}
V_r={l\over \cos(\psi^{*}_n)\arccos(1-8bl/(a\cos(\psi^{*}_n)))}~.
\label{eq5}
\end{equation}

It is always good to remember that, this resonant phenomenon only occurs when the defocusing mechanism \cite{ref25} in this billiard is not working, according to the expression $l/2R \approx 4bl/a^2 > 1$.\

\begin{figure}[h!]
\centerline{\includegraphics[width=0.6\linewidth]{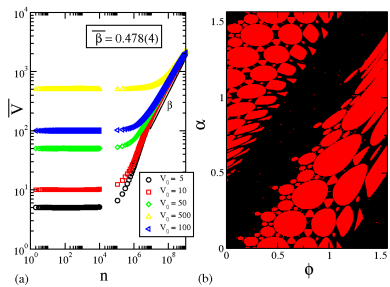}}
\caption{(color online). (a) Evolution of some curves of $\overline{V}$ as function of the number of collisions for $V_0>V_r$. The control parameters used were $a=0.5$, $b=0.01$, $l=1$, $B_0=0.01$, which give us $V_r=1.2$. (b) Grade of $500\times500$ initial conditions $(\alpha,\varphi \in [0,\pi/2])$ for $V_0=5$.} 
\label{fig3} 
\end{figure}

\begin{figure}[h!]
\centerline{\includegraphics[width=0.6\linewidth]{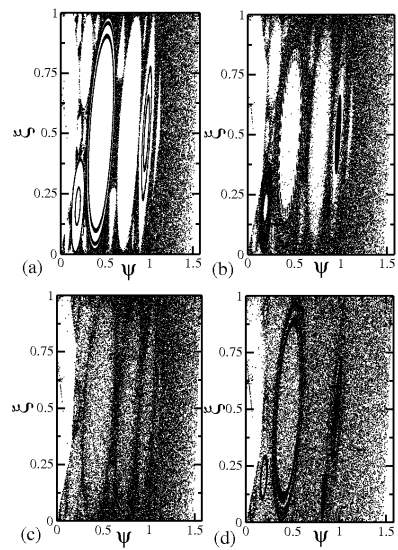}}
\caption{(color online) Phase Space for 25 different initial conditions iterated up to $10^6$ collisions. In (a) $V_0=5$, (b) $V_0=1.3$, (c) $V_0=1.2$ and (d) $V_0=0.5$.} 
\label{fig4} 
\end{figure}

If the initial velocity of the particle is higher than the resonant one, we observe typical FA behavior when we study the evolution of $\overline{V}$ as function of the number of collisions. Figure \ref{fig3}, shows some evolutions of $\overline{V}$ for 5000 different initial conditions $(\alpha,\varphi)$, evaluated up to $10^9$ collisions, for $V_0>V_r$, where $\beta\approx0.5$ is the growth exponent. However, not all initial conditions are accelerated. In the dark part of Fig.\ref{fig3}(b) are the initial conditions who suffered the effects of FA. On the other hand, the initial conditions that are inside the islands, do not experience this unlimited energy growth. We evaluated each pair $(\alpha,\varphi)$ up to $10^8$. If after this time, the particle has its velocity increased at least one order of magnitude, we considered that the particle suffered FA. As far we can study, the initial conditions who not suffered FA, are in the regular region of the phase space shown in Fig.\ref{fig4}(a). It is important to clarify, that the control parameters used in Fig.\ref{fig3}, will be the same used in all figures of this letter.\

However, if the initial velocity has a smaller value than the resonant critical one given by Eq.(\ref{eq5}), the particle experiences a resonance, and stays for limited time, surrounding the stability regions. In this time interval, the particle can penetrate into the neighborhood of fixed points, as a result, the whole region of the phase space became accessible \cite{ref24}, as shown Fig.\ref{fig4},where some phase spaces are shown for different values of $V_0$. It is clearly to see, that if $V_0>V_r$, the particle does not experience the resonance and the phase space does not become all accessible, differently in what happen if $V_0\leq V_r$.\

\begin{figure}[h!]
\centerline{\includegraphics[width=0.6\linewidth]{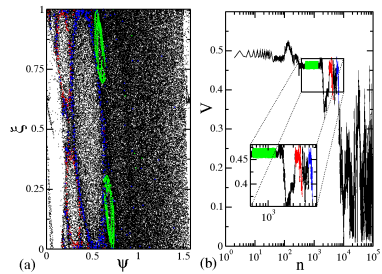}}
\caption{(color online) In (a) we show an orbit with an initial stickiness and trajectories penetrating through the stability islands for $V_0=0.5$; and in (b) Velocity evolution as function of $n$, for the same initial condition of (a), where the colored parts, corresponds to the same stickiness behavior shown in (a).} 
\label{fig5} 
\end{figure}

This resonance can appear in systems with mixed phase space \cite{ref25} and it causes a stickiness phenomenon \cite{ref26,ref27}, where some orbits can be trapped, surrounding stability islands for a limited time interval. We do believe, that this orbits in stickiness regime, are the responsible for the deceleration mechanism of the velocity. Looking at Fig.\ref{fig5}(a,b) where a single initial condition is iterated up to $10^5$ collisions, one can see, that after the orbit experiences a initial stickiness regime (colored parts in both Figs.\ref{fig5}(a,b)), the velocity has a decrease.

\begin{figure}[h!]
\centerline{\includegraphics[width=0.6\linewidth]{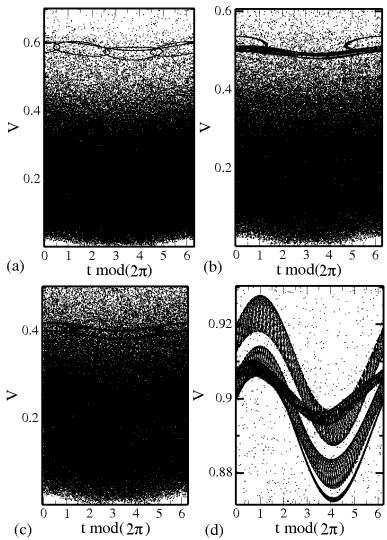}}
\caption{(color online) Phase Space in the coordinates $V$ and $t~~mod(2\pi)$ for 25 different initial conditions iterated up to $10^6$ collisions. In (a) $V_0=0.6$, (b) $V_0=0.5$, (c) $V_0=0.4$ and (d) a resonance zoom-in with $V_0=0.9$.} 
\label{fig6} 
\end{figure}

This kind of stickiness influence is interesting and need more attention, since this kind of dynamics is considered an open problem \cite{ref28}. We can see better this effect in Fig.\ref{fig6}(a,b,c) where we show the dynamics evolution in a different phase space where the coordinates are velocity and the time taken $mod(2\pi)$ for 50 initial conditions iterated up to $10^5$. We can see some darker lines in the region of the initial velocity, indicating the orbits in stickiness regime in the range of the initial velocity. A zoom-in over these dark curves is shown in Fig.\ref{fig6}(d), where we are able to see the rotations, during this initial stickiness behavior.\

When we consider the average velocity, we will found similar behavior as shown Fig.\ref{fig5}. The $\overline{V}$ curves experiences a constant velocity initial dynamics, where the orbits are experiencing this stickiness behavior confirmed by Figs.\ref{fig5} and \ref{fig6}; and then they start decreasing and reaches a constant plateau, remaining in this dynamics for long periods (in our numerical simulations up to $3\times10^6$).\

\begin{figure}[h!]
\centerline{\includegraphics[width=0.6\linewidth]{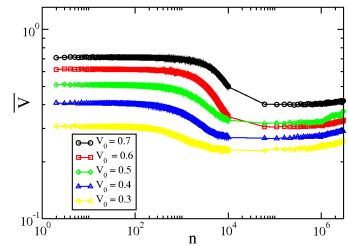}}
\caption{(color online) Behavior of the average velocity over an ensemble of 5000 different initial conditions, for some different values of initial velocity. The velocity decreasing is evident.} 
\label{fig7} 
\end{figure}

To conclude, we have presented that orbits in stickiness regime can act as a deceleration mechanism for FA in a time-dependent stadium billiard,even with no dissipation at all. At principle, we expect that this mechanism can be extended to other time-dependent billiards.\

ALPL and AL acknowledges FAPESP for financial support and E.D.L. acknowledges CNPq, FAPESP, CAPES and FUNDUNESP.

\end{document}